\documentclass[11pt]{article}
\usepackage{graphicx}
\usepackage{cite}

\begin{document}

%
%
\title{\Large\bf
Transverse single spin asymmetry in direct photon production in
polarized pA collisions }

\author{Andreas Sch\"{a}fer and Jian Zhou
 \\[0.3cm]
{\normalsize\it Institut f\"{u}r Theoretische Physik,Universit\"{a}t
Regensburg, Regensburg, Germany}}

\maketitle

\begin{abstract}
\noindent  We study the transverse single spin asymmetry in direct
photon production in pA collisions with incoming protons being
transversely polarized. To facilitate the calculation, we formulate
a hybrid approach in which  the nucleus is treated in the Color
Glass Condensate (CGC) framework while the collinear twist-3
formalism is applied on the proton side. It has been found that an
additional term which arises from color entanglement shows up in the
spin dependent differential cross section. The fact that this
additional term is perturbatively calculable allows us to
quantitatively  study color entanglement effects.
\end{abstract}

\section{Introduction}
The phenomenology of transverse single spin asymmetries (SSAs) in
high energy scattering has attracted a lot of attention and has been
under intense investigation during the past few decades. The large
size of the observed SSAs for single inclusive hadron production
came as a big surprise and can not be understood in the naive parton
model~\cite{Kane:1978nd,Ma:2008gm}.
 It signals that QCD phenomena are in general much richer than in the well studied collinear limit. This opens up many
possibilities. For example, SSAs could be especially sensitive to
saturation phenomena. However, QCD beyond the collinear limit is
usually more difficult and less under theoretical control, such that
presently one is still searching for the most adequate theoretical
frameworks to treat them. This contribution constitutes one more
attempt along these lines.

Various recent  studies show that it is necessary to take into
account initial/final state gluon re-scattering interactions in
order to generate large SSAs. Two ways of incorporating these
initial/final state interactions are based on the transverse
momentum dependent(TMD) factorization~\cite{Sivers:1990fh,
Collins:1992kk}
 and the collinear twist-3 factorization~\cite{Efremov:1981sh,Qiu:1991pp,Qiu:1991wg,Ji:1992eu,Yuan:2009dw,Beppu:2010qn}, respectively.
In TMD factorization, the naive time reversal odd TMD distributions
and fragmentation function, known as the quark/gluon Sivers
functions~\cite{Sivers:1990fh} and the Collins fragmentation
function~\cite{Collins:1992kk}
 can account for the large SSAs, while in the collinear twist-3 approach,
 SSAs can arise from a twist-3 quark gluon correlator, the
so-called Efremov-Teryaev-Qiu-Sterman(ETQS) function
~\cite{Efremov:1981sh,Qiu:1991pp}, a tri-gluon correlation
functions~\cite{Ji:1992eu,Beppu:2010qn}, and a twist-3 collinear
fragmentation functions~\cite{Yuan:2009dw}. It has been established
that the $k_\perp$ moment of the Sivers function and the Collins
function can be related to the ETQS function and the corresponding
twist-3 collinear fragmentation functions,
respectively~\cite{Boer:2003cm,Yuan:2009dw}.

SSAs observed in various processes like  pion production in single
polarized pp collisions $p^{\uparrow} p\rightarrow \pi X$ or in
SIDIS $e p^{\uparrow} \rightarrow \pi X$ receive contributions from
both sources: the Sivers mechanism and the Collins mechanism. A very
recent study has shown that the Collins effect described within the
collinear twist-3 framework might be the dominant contribution to
the spin asymmetry in polarized pp
collisions~\cite{Kanazawa:2014dca}. Nevertheless, it  would  be
crucial to unambiguously pin down the Sivers mechanism in polarized
pp collisions, as the so-called "sign mismatch"
problem~\cite{Kang:2011hk} (for a different version of this problem,
see also~\cite{Metz:2012ui}) is still unsolved.  It is thus
desirable to investigate SSAs for cases of particle production in
polarized $pp$ collisions for which the Collins effect is absent.
Possible options are the SSA in direct photon
production~\cite{Qiu:1991wg,Kouvaris:2006zy,Kanazawa:2012kt}, jet
production~\cite{Gamberg:2012iq,Kanazawa:2012kt,Nogach:2012sh},
 or  heavy quarkonium production in polarized pp collisions~\cite{Yuan:2008vn,Schafer:2013wca}.
 The SSA for direct photon production in polarized pp collisions
 has been calculated in the collinear twist-3 approach in~\cite{Qiu:1991wg,Kouvaris:2006zy,Kanazawa:2012kt}.
 In the present work, we extend this analysis to single polarized p$^{\uparrow}$A collisions.

Though most work in this field focuses on SSAs in ep$^{\uparrow}$ or
pp$^{\uparrow}$ collisions, there exist a few exploratory
investigations devoted to the study of SSAs in p$^{\uparrow}$A
collisions. The authors of Ref.~\cite{Boer:2006rj} investigated the
SSA for inclusive pion production at forward rapidities in
p$^{\uparrow}$p collisions using a hybrid approach in which the
target proton is treated in the CGC framework~\cite{McLerran:1993ni}
while the spin-transverse momentum correlation in the projectile
proton is described by the Sivers distribution. Their analysis can
be straightforwardly applied to p$^{\uparrow}$A collisions.
Following the same line of reasoning, the SSA of Drell-Yan lepton
pairs produced in p$^{\uparrow}$A collisions was computed
in~\cite{Kang:2012vm}. On the other hand, the SSA for inclusive pion
production caused by the Collins mechanism after the transversely
polarized quark from the projectiles is scattered off the background
gluon field of the nucleus  was investigated in
Ref.~\cite{Kang:2011ni}. Furthermore, a recent GCG calculation
suggests that SSAs also can be generated by the interaction of the
spin-dependent light-cone wave function of the projectile with the
target gluon field via C-odd odderon
exchange~\cite{Kovchegov:2012ga}.

The purpose of the present work is to study SSA for direct photon
production in p$^{\uparrow}$A collisions and to decide whether it
provides a sensitive tool to establish and study saturation effects.
First one can note that the contribution from fragmentation to the
spin asymmetry for direct photon production is found to be
negligible~\cite{Gamberg:2012iq}. Next one observes that TMD
factorization can not be applied on proton side for lack of an
additional hard scale. Moreover, it has been shown that the odderon
exchange does not give rise to a SSA for direct photon
production~\cite{Kovchegov:2012ga}. Therefore, the only possible
source for a sizeable SSA is the Sivers effect which can only  be
described within the collinear twist-3 approach for the process
under consideration.
 To do so, we formulate a novel hybrid approach in which the nucleus is treated in the
Color Glass Condensate (CGC) framework while the collinear twist-3
formalism is applied on the proton side. In this hybrid approach, we
take into account one extra gluon exchange from the proton side and
sum gluon re-scattering to all orders on the nucleus side.

The resulting spin dependent differential cross section computed in
this hybrid approach is proportional to a convolution of the ETQS
function and various Wilson lines. These Wilson lines can be further
related to two different types of $k_\perp$ dependent gluon
distributions, one of which is the dipole type gluon TMD. The other
arises from a color entanglement effect which is due to the
 non-trivial interplay of gluons from both the nucleon and nucleus~\cite{Rogers:2010dm}.
 This effect seems to be a unique feature of non-abelian theories, i.e. it is linked to one of the
most fundamental aspects of QCD.
 The fact that this additional term can be perturbatively calculated in the Mclerran-Venugopalan~(MV)
model~\cite{McLerran:1993ni} allows us to quantitatively  study the
color entanglement effect and to test the MV model. A measurement of
this observable would also provide a hint  to the size of
generalized TMD factorization breaking effect. Let us note that it
was argued that such a color entanglement effect could  also
manifest itself through azimuthal asymmetries in the Drell-Yan
process~\cite{Buffing:2013dxa}.

In a more general context, the present work is part of the effort to
address
 the interplay between spin physics and saturation physics.
Apart from the studies mentioned above, early work in this very
active field includes the study of small x evolution of spin
dependent structure function $g_1$~\cite{Bartels:1995iu} and of the
quark Boer-Mulders distribution and the linearly polarized gluon
distribution
 inside a large nucleus, see Refs.~\cite{Metz:2011wb}.
The small x evolution equations for the linearly polarized gluon
distributions were derived in Ref.~\cite{Dominguez:2011br}. Several
ways of accessing the linearly polarized gluon distributions inside
a large nucleus have been discussed
in~\cite{Metz:2011wb,Dominguez:2011br,Schafer:2012yx,Akcakaya:2012si}.
Furthermore, the asymptotic behavior of transverse single spin
 asymmetries at small x was discussed in Ref.\cite{Schafer:2013opa,Zhou:2013gsa}.
It has been shown that SSAs at small x are generated by polarized
odderon exchange whose size is determined by the anomalous magnetic
moment~\cite{Zhou:2013gsa}. The quark Sivers function was computed
in the quasi-classical Glauber-Mueller/MV
approximation~\cite{Kovchegov:2013cva}. More recently, the authors
of the paper~\cite{Altinoluk:2014oxa} have investigated the spin
asymmetries in pA collisions by going beyond the Eikonal
approximation.

The paper is organized as follows.  In section II, we briefly review
the existing calculations for direct photon production, including
the collinear twist-3 calculation for direct photon production in
polarized p$^{\uparrow}$p collisions and the CGC calculation for
direct photon production in unpolarized pA collisions. In section
III, we develop the hybrid approach and explain all technical steps
in details. We focus on the derivative term contribution and
identify a term arising from the color entanglement effect. It is
shown that the spin dependent differential cross section derived in
collinear factorization can be recovered from our result without
 color entanglement effect being incorporated in the kinematical
limit where the produced photon transverse momentum is much larger
than the saturation scale. The paper is summarized in section IV.

\section{ Brief review of existed calculations for direct photon production}
In this section, we review how the calculation of the SSA for direct
photon production is formulated within the collinear twist-3
approach in p$^{\uparrow}$p collisions, following by a brief
reminder of the application of the CGC framework to direct photon
production in unpolarized pA collisions.

\subsection{SSA in direct photon production in  p$^{\uparrow}$p collisions }
The dominant production mechanism for prompt photons in high energy
collisions is Compton scattering $gq \rightarrow\gamma q$. We start
by introducing the relevant kinematical variables and assign
4-momenta to the particles according to
\begin{eqnarray}
 g(x_g' \bar P )\ + \ q(xP)\longrightarrow \gamma(l_\gamma) \  + \ q(l_q)
\end{eqnarray}
where $\bar P^\mu=\bar P^- n^\mu$ and  $ P^\mu= P^+ p^\mu$ with
$n^\mu$ and $p^\mu$ being the commonly defined light cone vectors,
normalized according to $p \cdot n=1$. The Mandelstam variables are
defined as: $S=(P+\bar P)^2$, $T=(P-l_q)^2$ and
$U=(P-l_\gamma)^2$.
 The corresponding unpolarized Born cross section reads,
\begin{eqnarray}
\frac{d^3  \sigma}{d^2 l_{\gamma\perp} dz }=
 \frac{\alpha_s \alpha_{em} }{N_c} \
\frac{z[1+(1-z)^2]}{l_{\gamma\perp}^4} \sum_q e_q^2 \int^1_{x_{min}}
dx \ f_q(x)  x_g'G(x_g') \label{colunp}
\end{eqnarray}
where $z\equiv l_\gamma \cdot n/(x P \cdot n) $ is the fraction of
the incoming quark momentum $xP$ carried by the outgoing photon,
 and $l_{\gamma\perp}$ is the photon transverse momentum.
The meaning of the other coefficients should be self-evident. Note
that $x_g'=\frac{-xT}{xS+U}$ is a function of $x$;
 and $x_{min}$ is given by $x_{min}=\frac{-U}{S+T}$.
In the above formula, $f_q(x)$ and $G(x_g')$ are the usual
integrated quark and gluon distributions, respectively.

\begin{figure}[t]
\begin{center}
\includegraphics[width=12cm]{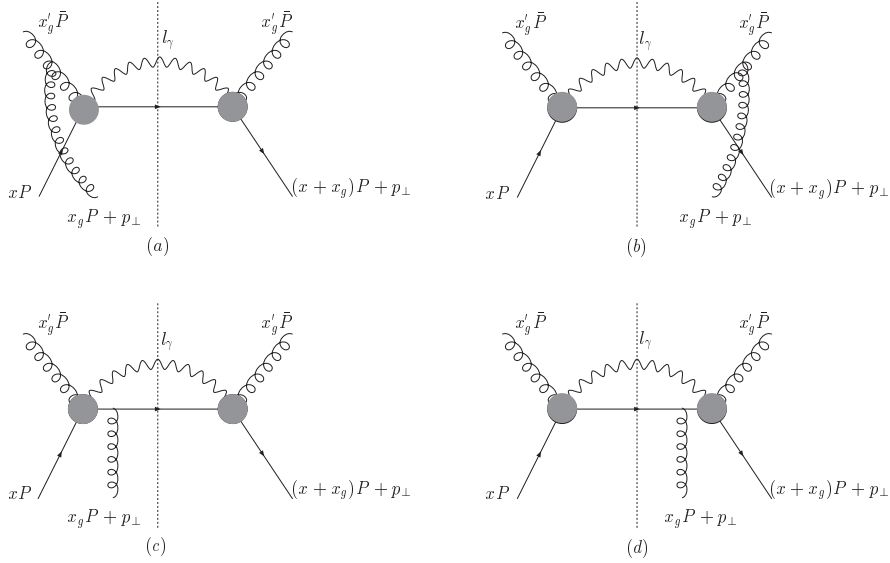}
\caption {Diagrams contributing to the single spin asymmetry for
direct photon production in p$^{\uparrow}$p collisions. Grey circles
indicate all possible photon line attachments.
 The mirror diagrams are not shown here.
The contributions from diagrams (c) and (d) to the spin asymmetry
cancel. } \label{1}
\end{center}
\end{figure}

To generate the spin asymmetry, one additional gluon must be
exchanged between the active partons and the remanent part of the
polarized proton projectile. The hard part, if an additional gluon
is attached, can be calculated perturbatively, while the
non-perturbative part describes the relevant three parton
correlations. The strong interaction phase factor necessary for
having a non-vanishing spin asymmetry arises from the interference
between an imaginary part of the partonic scattering amplitude with
an extra gluon and the real scattering amplitude without a gluon
attachment, as shown in Fig.1. The imaginary part is due to the pole
of the parton propagator associated with the integration over the
gluon momentum fraction $x_g $.  This effectively implies that one
of the internal parton lines goes on shell. To isolate the imaginary
part of such poles, the identity of distributions: $\frac{1}{x_g \pm
i\epsilon}= {\rm PV} \frac{1}{x_g}\mp i\pi \delta(x_g)$ was used.
Depending on which propagator's pole contributes, the amplitude may
get contributions from $ x_g=0$ (``soft-pole") and $x_g\neq 0$
(``hard-pole" ).

It is convenient to carry out the calculation in the covariant
gauge, in which the leading contribution of the exchanged gluon is
the "plus" component $A^+$. The gluon's momentum is given by
$p_g=x_gP+p_{\perp}$, where $x_g$ is the longitudinal momentum
fraction with respect to the polarized proton. In order to calculate
consistently with twist-3 accuracy, one has to expand the hard part
in the gluon transverse momentum,
\begin{eqnarray}
H(x_gP+ p_{\perp},l_\gamma)=H(x_gP,l_\gamma)+ \frac{\partial H(x_gP+
p_{\perp},l_\gamma)}{\partial p_\perp^\rho}|_{ p_\perp=0}
 \  p_\perp^\rho+ ...
\end{eqnarray}
In the above formula, the first term only contributes to the
unpolarized Born cross section. We thus have to keep the linear term
in $p_\perp$ at twist-3 level. In the second term, the $p_\perp$
factor can be combined with $A^+$ to yield $\partial^\perp A^+$,
which is an element of the field strength tensor $F^{\partial +}$.
The above expansion allows us to integrate over three of the four
components of each of the loop momenta $p_g$. The four-dimensional
integral is reduced to a convolution in the light-cone momentum
fractions of the initial partons. At this step, the relevant three
parton correlation can be cast into the form of the ETQS function
defined as~\cite{Efremov:1981sh,Qiu:1991pp},
\begin{eqnarray}
T_{F,q}(x_1,x_2)&=&\int \frac{dy_1^-dy_2^-}{4 \pi} e^{ix_1P^+ y_1^-
+ i(x_2-x_1)P^+y^-_2}
\nonumber \\
&& \times \langle P,S_\perp | \bar{\psi}_q(0)\gamma^+ g
\epsilon^{S_\perp \sigma n p} F_ \sigma^{ \ +}(y_2^-)\psi_q(y_1^-) |
P,S_\perp \rangle
\end{eqnarray}
where we have suppressed Wilson lines. $S_\perp$ denotes the proton
transverse spin vector. Note that our definition of the ETQS
functions differs by a factor $ g$ from the convention used in
Ref.~\cite{Kouvaris:2006zy}. This ETQS function plays an important
role in describing SSA phenomenology.

Making use of the ingredients described above, the calculation is
straightforward. The spin dependent cross section has been
calculated and given in~\cite{Qiu:1991wg,Kouvaris:2006zy},
\begin{eqnarray}
 \frac{d^3 \Delta \sigma}{d^2 l_{\gamma\perp}dz}&=&
 \frac{\alpha_s \alpha_{em} N_c }{N_c^2-1} \ \frac{z[1+(1-z)^2]}{l_{\gamma\perp}^4}(z-1)
\left ( \frac{\epsilon^{l_{\gamma} S_\perp  n p}}{
l_{\gamma\perp}^2} \right )
\nonumber \\
&& \times \sum_q e_q^2 \int^1_{x_{min}} dx \  x_g'G(x_g')
\left [ T_{F,q}(x,x)-x \left (\frac{d}{dx} T_{F,q}(x,x) \right )
\right ]
\end{eqnarray}
where we have omitted the  soft fermion pole
contribution~\cite{Kanazawa:2012kt}. In the next section, we will
show that the above spin dependent cross section can be recovered
from the proposed hybrid approach in the kinematical limit where the
saturation scale is much smaller than the produced photon transverse
momentum after neglecting  terms arising from color entanglement effect.

\subsection{Photon production in unpolarized pA collisions}
We now move on to review the CGC calculation for direct photon
production in unpolarized pA collisions which has been done in
Ref.~\cite{Gelis:2002ki}. Roughly speaking, the CGC calculation for
this process differs from the collinear factorization calculation in
two ways. Firstly, in the small x region,
  transverse momenta carried by gluons are not necessarily much smaller than their longitudinal momenta.
One thus should keep gluon transverse momenta when computing the
hard part. We fix kinematical variables accordingly,
\begin{eqnarray}
 g(x_g' \bar P+k_\perp )\ + \ q(xP)\longrightarrow \gamma(l_\gamma) \  + \ q(l_q) \ .
\end{eqnarray}
Secondly, due to the high gluon number density at small x, it is
necessary to resum gluon re-scattering to all orders.

\begin{figure}[t]
\begin{center}
\includegraphics[width=14cm]{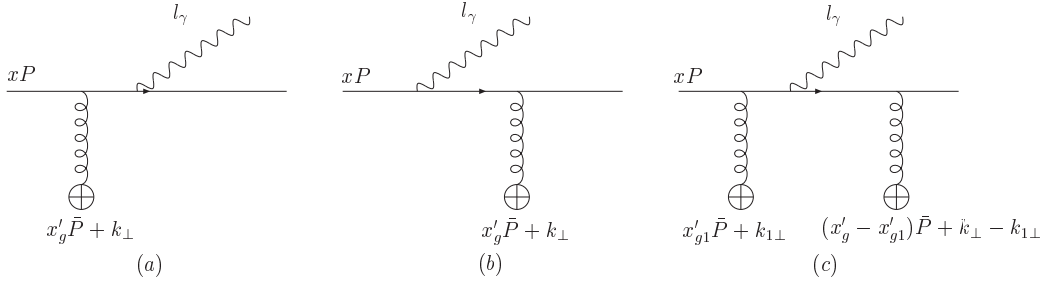}
\caption[] {Diagrams contributing to the unpolarized cross section
for direct photon production in pA collisions. The gluon line
terminated by a cross surrounded with a circle denotes a  classical
field $A_A$ insertion. The contribution from diagram(c) vanishes
because two poles are lying on the same half plane.} \label{2}
\end{center}
\end{figure}

The multiple scattering between  quark  and the classical color
field of the nucleus can be readily resummed to all
orders~\cite{Balitsky:1995ub,McLerran:1998nk}. This gives rise to a
path-ordered gauge factor along the straight line that extends in
$x^+$ from minus infinity to plus infinity. More precisely, for a
quark with incoming momentum
 $l$ and outgoing momentum  $l+k$, the path-ordered gauge factor reads,
\begin{equation}
 2 \pi  \delta(k^+)
 n^\mu [U(k_\perp)-(2\pi)^2 \delta(k_\perp)] \,,
\end{equation}
with
\begin{equation}
U(k_\perp) =\int d^2 x_\perp e^{ i  k_\perp \cdot  x_\perp}
U(x_\perp)\,,
\end{equation}
and
\begin{equation}
U(x_\perp)= {\cal P} e^{ig\int_{-\infty}^{+\infty} dx^+ A^-_A (x^+,
\ x_\perp)\cdot t }  \ ,
\end{equation}
where  $t$ is the generators in the fundamental representation. We
use this as building block to compute the amplitude for direct
photon production in high energy $pA$ collisions. It is
straightforward to obtain the production amplitude for diagram (a)
illustrated in Fig.~\ref{2},
\begin{equation}
{\cal M}_{2a}=\bar u(l_q) (ie) \varepsilon \!\!\!/ S_F(xP+x_g'\bar
P+k_\perp)n \!\!\!/ u(xP) \left [U(k_\perp)-(2\pi)^2 \delta(k_\perp)
\right ]
\end{equation}
where a delta function is suppressed. In the above formula,
$\varepsilon^\mu$ is the polarization vector of the produced photon,
 and $S_F(xP+x_g'\bar P+k_\perp)=
i\frac{xP\!\!\!\!/+x_g'\bar P\!\!\!\!/+k_\perp
\!\!\!\!\!\!\!/}{(xP+x_g'\bar P+k_\perp)^2+i\epsilon}$ is the quark
propagator.  The contribution of diagram (b) in Fig.~\ref{2} to the
amplitude is similarly given by,
\begin{equation}
{\cal M}_{2b}=\bar u(l_q) n \!\!\!/ S_F(xP-l_\gamma) (ie)
\varepsilon \!\!\!/  u(xP) \left [U(k_\perp)-(2\pi)^2
\delta(k_\perp) \right ]
\end{equation}
The contribution of diagram (c) in Fig.~\ref{2} vanishes. This is so
because both $x_{g1}'$ poles lie below the real axis, such that one
can close the integration contour above the real axis and get a
vanishing contribution. The total amplitude is thus given by ${\cal
M}_2={\cal M}_{2a}+{\cal M}_{2b}$.
 By squaring the amplitude, one obtains the cross section~\cite{Gelis:2002ki},
\begin{eqnarray}
\frac{d \sigma}{d^2 l_{\gamma\perp} dz} =
 \frac{\alpha_{em} \alpha_s }{N_c} \frac{1}{l_{\gamma\perp}^2}
\frac{1+(1-z)^2}{z} \sum_q e_q^2 \int_{x_{min}}^1 dx \int
\frac{  d^2 k_\perp}{(k_\perp-l_{\gamma\perp}/z)^2}
x_g'G_{DP}(x_g',k_\perp)f_q(x) \label{CGCunp}
\end{eqnarray}
where $x_g'G_{DP}(x_g',k_\perp)$ is the dipole type gluon TMD,
defined as
\begin{eqnarray}
x_g' G_{DP}(x_g', k_\perp)=\frac{k_\perp^2 N_c}{2 \pi^2 \alpha_s}
\int \frac{d^2x_\perp d^2y_\perp }{(2\pi)^2} e^{i  k_\perp \cdot
(y_\perp-x_\perp)} \frac{1}{N_c} \langle {\rm Tr} \left [
U(x_\perp)U^\dag(y_\perp)  \right ] \rangle_{x_g'}
\end{eqnarray}
The Wilson lines appearing in the above formula can be explicitly
evaluated in the MV model~\cite{McLerran:1993ni}. The resulting
dipole gluon distribution reads
\begin{eqnarray}
x_g' G_{DP}(x_g', k_\perp)=\frac{k_\perp^2 N_c}{2 \pi^2 \alpha_s}
 \pi R_0^2 \int \frac{d^2r_\perp }{(2\pi)^2} e^{i  k_\perp \cdot r_\perp}
e^{-\frac{1}{4} r_\perp^2 Q_{sq}^2}
\end{eqnarray}
where $Q_{sq}^2 = \alpha_s C_F \mu {\rm ln} \frac{1}{r_\perp^2
\Lambda_{QCD}^2}$ is the quark saturation momentum with $\mu$ being
the transverse color source density for a nucleus. Here $R_0$ is the
radius of nucleus.

In the large transverse momentum region $l_{\gamma\perp}^2 \gg
Q_{sq}^2 \sim   k_\perp^2  $, the denominator in Eq.~(\ref{CGCunp})
can be approximated as: $1/(k_\perp-l_{\gamma\perp}/z)^2 \approx
z^2/l_{\gamma\perp}^2 $.
 After making this approximation and using the relation
\begin{eqnarray}
\int d^2 k_\perp x_g' G_{DP}(x_g', k_\perp)=x_g' G(x_g') \ ,
\label{relation1}
\end{eqnarray}
  one is able to reproduce Eq.~(\ref{colunp}) which was obtained from collinear factorization.

\section{SSA for photon production in p$^{\uparrow}$A collisions}
\begin{figure}[t]
\begin{center}
\includegraphics[width=12cm]{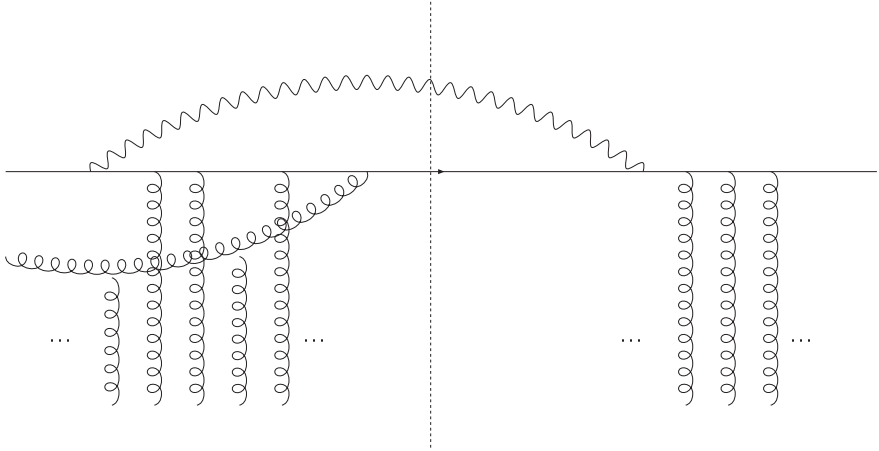}
\caption[] {A typical diagram contributing to the SSA in direct
photon production in polarized pA collisions. The multiple
re-scattering of the incoming partons (including the unpolarized
quark and longitudinally polarized gluon) from the proton off the
classical gluon field of the nucleus needs to be resummed to all
orders. } \label{3}
\end{center}
\end{figure}

To calculate the SSA for direct photon production in polarized pA
collisions, we have to take into account one extra gluon exchange
between the active partons and the remanent part of the polarized
proton,  while gluon re-scattering inside the nucleus must be
resummed to all orders. A typical diagram contributing to this
process is illustrated in Fig.~\ref{3}. It is worthwhile to mention
that gluons from the nucleus could also interact with the color
source inside the proton.  Such an interaction is not shown in
Fig.~\ref{3}.
 In this section, we derive the spin dependent amplitude in the CGC framework.  We further
calculate the derivative term contribution with the obtained
amplitude, and also show that the full polarized cross section can
be reduced to the one computed from the standard collinear twist-3
approach at high photon transverse momentum provided that the $1/N_c^2$ suppressed color
entanglement effect has been neglected.

\subsection{Derivation of the spin dependent amplitude}
As mentioned in the previous section, the multiple scattering
between the incoming quark and the classical color field of the
nucleus can be resummed into a Wilson line. Similarly, this
procedure also applies to the case for which the incoming parton is
a transversely polarized gluon. However, in the process under
consideration, multiple scattering between incoming gluon and
background gluon field of the nucleus can not be described by a
simple Wilson line, since the incoming gluon from the proton side is
longitudinally polarized.

The formula for a longitudinally  polarized gluon scattering off a
nucleus has been worked out in Ref.~\cite{Blaizot:2004wu}. The
expression for  the gauge field created through the fusion of the
incoming gluon from the proton and small x gluons from the nucleus
contains
 both singular terms (proportional to $\delta(x^+)$) and regular terms,
\begin{eqnarray}
A^\mu(q)=A^\mu_{reg}(q)+\delta^{\mu-} A^-_{sing}(q) \ .
\end{eqnarray}
The regular terms $A^\mu_{reg}$ are given by
\begin{eqnarray}
A^\mu_{reg}&= & A_p^\mu \nonumber \\ &+& \frac{ig}{q^2+iq^+\epsilon}
\int \frac{d^2 p_\perp}{(2\pi)^2} \left \{ C_U^\mu(q,p_\perp) \left
[ \tilde U(k_\perp)-(2\pi)^2 \delta(k_\perp) \right ]  \right .\
\nonumber \\ && \ \ \ \ \ \ \ \ \ \ \ \ \ \  + \left .\
C^\mu_{V,reg}(q) \left [ \tilde V(k_\perp)-(2\pi)^2 \delta(k_\perp)
\right ] \right \} \frac{\rho_p(p_\perp)}{p_\perp^2} \label{dressed}
\end{eqnarray}
where $\rho_p(p_\perp)$ is the color source distribution inside a
proton,
 and $A_p^\mu$ is the gauge field created by the proton alone. In the MV model, it is given by,
\begin{eqnarray}
A^\mu_p=2 \pi g \delta^{\mu +} \delta(q^-)
\frac{\rho_p(q_\perp)}{q_\perp^2} \label{proton} \ ,
\end{eqnarray}
In second term of the formula \ref{dressed}, $p_\perp$ is the
momentum carried by the incoming gluon from the proton and $k_\perp$
defined as $k_\perp=q_\perp-p_\perp$ is the momentum coming from the
nucleus. For the polarized case, there exists a correlation between
the transverse momentum $p_\perp$ and the transverse proton spin
vector $S_\perp$. As shown below, such a correlation can be
described by the ETQS function, and leads to a SSA for direct photon
production.
 The four vectors $C_U^\mu(q,p_\perp)$ and $C_{V,reg}^\mu$ are given by the following relations
\begin{eqnarray}
&& C^+_U(q,p_\perp)=-\frac{{ p}_\perp^2}{q^-+i\epsilon}, \ \
C_U^-(q,p_\perp)=\frac{{k}_\perp^2-{q}_\perp^2}{q^++i\epsilon}, \ \
C^i_U(q,p_\perp)=-2{  \rm p}_\perp^i
 \\
 &&
C_{V,reg}^\mu(q)=2q^\mu-\delta^{-\mu} \frac{q^2}{q^++i\epsilon}
\end{eqnarray}
where the subscript $'reg'$ indicates that the corresponding term of
$A^\mu$ does not contain any $\delta(x^+)$ when expressed in
coordinate space. Here, we specified the $q^+$ pole structure
according to the fact that this term arises from an initial state
interaction. It is crucial to keep the imaginary part of this pole
in order to generate the non-vanishing spin asymmetry. The notation
${\rm p}_{\perp}$ is used to denote four dimension vector
 with  $p_{\perp}^2=-{\rm p}_\perp^2$.
$\tilde U(k_\perp)$ and $\tilde V(k_\perp)$ are the Fourier
transform of Wilson lines in the adjoint representation,
\begin{eqnarray}
\tilde U(k_\perp)=\int d^2 x_\perp e^{ik_\perp \cdot x_\perp} \tilde
U(x_\perp) , \ \ \ \tilde V(k_\perp)=\int d^2 x_\perp e^{ik_\perp
\cdot x_\perp} \tilde V(x_\perp)
\end{eqnarray}
with
\begin{eqnarray}
 \tilde U(x_\perp)&=& {\cal P} {\rm exp} \left [ ig \int_{-\infty}^{+\infty} dz^+ A_A^-(z^+,x_\perp) \cdot T \right ]  ,
 \\
 \tilde V(x_\perp)&=&{\cal P} {\rm exp} \left [ i\frac{g}{2} \int_{-\infty}^{+\infty} dz^+ A_A^-(z^+,x_\perp) \cdot T \right ]
\end{eqnarray}
where the $T$ are the generators of the adjoint representation. The
singular terms reads,
\begin{eqnarray}
A^-_{sing}(q)=-\frac{ig}{q^++i\epsilon} \int \frac{d^2
p_\perp}{(2\pi)^2} \left [ \tilde V(k_\perp)-(2 \pi)^2
\delta(k_\perp) \right ] \frac{\rho_p(p_\perp)}{p_\perp^2}
\end{eqnarray}

The peculiar Wilson line $\tilde V$ differs from the normal one
$\tilde U$ by a factor $1/2$ in the exponent. It has been
demonstrated that all terms containing $\tilde V$ cancel in the
unpolarized amplitudes for gluon production and quark pair
production in pA collisions~\cite{Blaizot:2004wu,Blaizot:2004wv}. It
will be shown below that the $\tilde V$ terms also drop out in the
spin dependent amplitude for direct photon production.
\begin{figure}[t]
\begin{center}
\includegraphics[width=12cm]{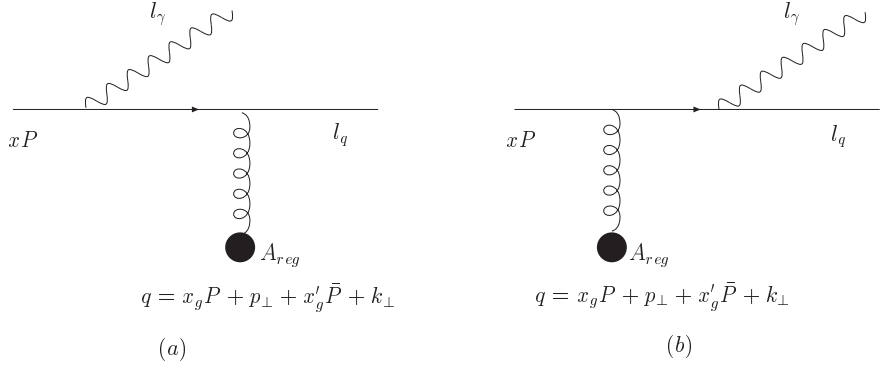}
\caption[] {The contribution from the regular terms to the spin
dependent amplitude. A black dot denotes a classical field $A_{reg}$
insertion.} \label{4}
\end{center}
\end{figure}

\begin{figure}[t]
\begin{center}
\includegraphics[width=14cm]{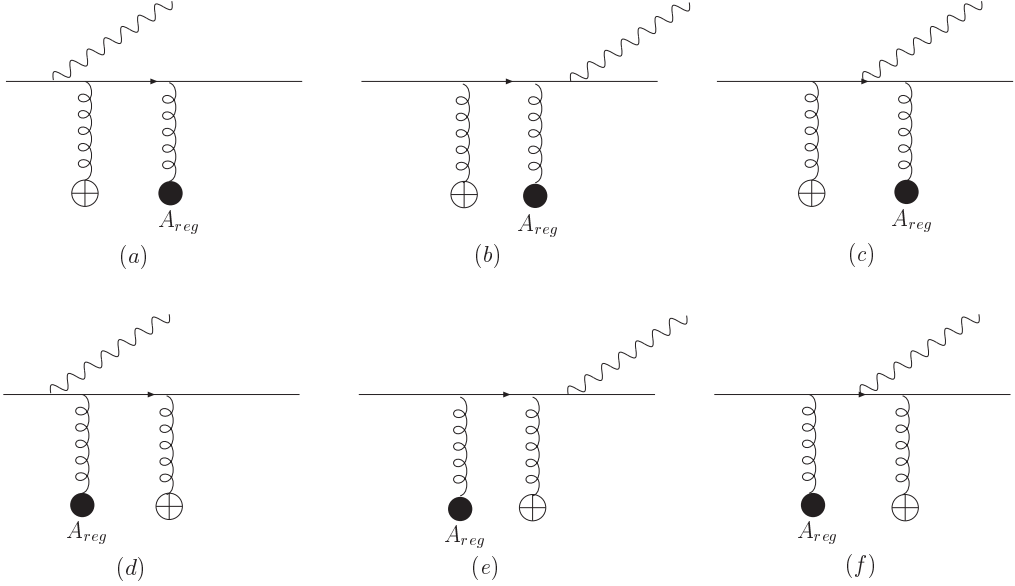}
\caption[] { The contribution from the regular terms to the spin
dependent amplitude. Black dot denotes a classical field $A_{reg}$
insertion, while the gluon line terminated by a cross surrounded
with a circle denotes a  classical field $A_A$ insertion.} \label{5}
\end{center}
\end{figure}

Following the method outlined in Ref.~\cite{Blaizot:2004wv}, we
calculate the contributions from the regular terms and the singular
terms separately.  Let us begin with the regular terms which do not
contain a delta function $\delta(x^+)$. Their contributions are
represented by the diagrams in Fig.~\ref{4} and Fig.~\ref{5}. The
amplitude from Fig.~(\ref{4}a) reads,
\begin{eqnarray}
{\cal M}_{4a}=[ig A^\mu_{reg}(q)] \left [ \bar u(l_q)\gamma_\mu t^a
S_F(xP-l_q) (ie) \varepsilon \!\!\! / u(xP) \right ]
\end{eqnarray}
where the momentum carried by the gluon produced through the fusion
of a longitudinally polarized gluon from the proton and a small x
gluons from the nucleus is given by $q=x_gP+p_\perp+x'_g \bar
P+k_\perp$. The soft gluon pole contribution to the amplitude ${\cal
M}_{4a}$  from the first term of Eq.~(\ref{dressed})  cancels
between the diagram Fig.~\ref{4}a and its mirror diagram in the same
way as between the diagrams Fig.~\ref{1}c and Fig.~\ref{1}d. In
addition, the contributions from the first part of $C_{V,reg}^\mu$
cancel between Fig.~\ref{4}a and Fig.~\ref{4}b due to the Ward
identity.  The spin dependent amplitude thus can be explicitly
written as,
\begin{eqnarray}
{\cal M}_{4a}&=&-ieg^2 \int \frac{d^2 p_\perp}{(2\pi)^2}
\frac{\rho_{p,a}(p_\perp)}{p_\perp^2} \nonumber \\ & \times&
  \bar u(l_q) \left \{ \frac{C_U \!\!\!\!\!\!\! / \ \ (q,p_\perp)}{q^2+i\epsilon} t^b
S_F(xP-l_q)  \varepsilon \!\!\! / \left [ \tilde U(k_\perp)-(2\pi)^2
\delta(k_\perp) \right ]_{ba}
 \right .\
 \nonumber \\
 &&  \ \ \ \ \ \ \ \ \ -
 \left .\  \frac{n \!\!\! / }{x_gP+i\epsilon} t^b
S_F(xP-l_q) \varepsilon \!\!\! / \left [ \tilde V(k_\perp)-(2\pi)^2
\delta(k_\perp) \right ]_{ba}
 \right \} u(xP)
\end{eqnarray}
 For the diagram in Fig.~\ref{4}b, it is easy to  verify
that the first term of the Eq.~(\ref{dressed}) gives rise to the
vanishing contribution. One thus obtains for the amplitude from
Fig.~\ref{4}b,
\begin{eqnarray}
{\cal M}_{4b}&=&-ie g^2 \int \frac{d^2 p_\perp}{(2\pi)^2}
\frac{\rho_{p,a}(p_\perp)}{p_\perp^2} \nonumber \\ & \times&
  \bar u(l_q) \left \{  \varepsilon \!\!\! /
  S_F(xP +q )
   \frac{C_U \!\!\!\!\!\!\! / \ \ (q,p_\perp) }{q^2+i\epsilon} t^b
\left [ \tilde U(k_\perp)-(2\pi)^2 \delta(k_\perp) \right ]_{ba}
 \right .\
 \nonumber \\
 &&  \ \ \ \ \ \ \ \ \ -
 \left .\  \varepsilon \!\!\! /
S_F(xP +q )  \frac{n \!\!\! / }{x_gP+i\epsilon} t^b \left [ \tilde
V(k_\perp)-(2\pi)^2 \delta(k_\perp) \right ]_{ba}
 \right \} u(xP)
\end{eqnarray}

The incoming quark also can directly interact with the classical
gluon field from the nucleus. This is illustrated in Fig.~\ref{5}.
The first term of $A_{reg}$ does not contribute to the spin
dependent part of the amplitudes from the diagrams in Fig.~\ref{5}.
One further notices that all of the $x_{g1}'$ poles in the
amplitudes from the diagrams in Fig.~\ref{5}d, Fig.~\ref{5}e and
Fig.~\ref{5}f are lying in the same half plane. Therefore, after
carrying out the $x_{g1}'$ integration using the theorem of the
residues, one has
\begin{eqnarray}
{\cal M}_{5d}={\cal M}_{5e}={\cal M}_{5f}=0
\end{eqnarray}
We are left with the contributions from Fig.~\ref{5}a, Fig.~\ref{5}b
and Fig.~\ref{5}c. After carrying out the $x_{g1}'$ integration, it
becomes evident that the contributions from the first part of the
$C_{V,reg}^\mu$ term cancel between the diagrams in Fig.~\ref{5}a,
Fig.~\ref{5}b and Fig.~\ref{5}c. With these simplifications, the
expression for the amplitude of Fig.~\ref{5}a is given by,
\begin{eqnarray}
{\cal M}_{5a}&=&\int \frac{d^4 k_1}{(2\pi)^4} 2\pi \delta(k_1^+) [ig
A^\mu_{reg}(q-k_1)]
 \nonumber \\ && \times
\bar u(l_q) \gamma_\mu t^a S_F(xP-l_\gamma+k_1) n\!\!\!/ \left [
U(k_{1\perp})-(2\pi)^2\delta(k_{1\perp}) \right ] S_F(xP-l_\gamma)
ie \varepsilon \!\!\!/ u(xP)
 \nonumber \\
&=&-ie g^2 \int \frac{d k^-_1 d^2 k_{1\perp} }{(2\pi)^3} \int
\frac{d^2 p_\perp}{(2\pi)^2}
 \frac{\rho_{p,a}(p_\perp)}{p_\perp^2} \ \bar u(l_q)
 \frac{C_U \!\!\!\!\!\!\!/  \ \ (q-k_1,p_\perp)}{(q-k_1)^2+i\epsilon}  t^b S_F(xP-l_\gamma+k_1) n\!\!\!/
 \nonumber \\ && \times
\left [ U(k_{1\perp})-(2\pi)^2\delta(k_{1\perp}) \right ]
S_F(xP-l_\gamma) \varepsilon \!\!\!/ u(xP) \left [ \tilde
U(k_\perp-k_{1\perp})-(2\pi)^2 \delta(k_\perp-k_{1\perp}) \right
]_{ba}
  \nonumber \\ &+&
 ie g^2 \int \frac{d^2 p_\perp}{(2\pi)^2}
 \frac{\rho_{p,a}(p_\perp)}{p_\perp^2} \int d^2x_\perp e^{ik_\perp \cdot x_\perp}
  \ \bar u(l_q)
 \frac{n \!\!\! / }{x_gP+i\epsilon} t^b  \left [ U(x_\perp)-1 \right ] S_F(xP-l_\gamma)
  \nonumber \\ && \times  \varepsilon \!\!\!/ u(xP)
\left [ \tilde V(x_\perp)-1 \right ]_{ba}
\end{eqnarray}
where we have applied  the Eikonal approximation to the quark
propagator $ S_F(xP-l_\gamma+k_1)$ which appears in the hard part
associated with the term containing  $\left [ \tilde V(x_\perp)-1
\right ]_{ba}$. The $k_1^-$ and $k_{1\perp}$ integrations have been
carried out in the second term after making the Eikonal
approximation. Following the similar procedure, we obtain the
amplitude from the diagram in Fig.~\ref{5}b,
\begin{eqnarray}
{\cal M}_{5b}&=& -ie g^2 \int \frac{d k^-_1 d^2 k_{1\perp}
}{(2\pi)^3} \int \frac{d^2 p_\perp}{(2\pi)^2}
 \frac{\rho_{p,a}(p_\perp)}{p_\perp^2} \ \bar u(l_q) \varepsilon \!\!\!/
S_F(xP+q)
 \frac{C_U \!\!\!\!\!\!\!/  \ \ (q-k_1,p_\perp)}{(q-k_1)^2+i\epsilon}  t^b S_F(xP+k_1)
 \nonumber \\ && \times  n\!\!\!/
\left [ U(k_{1\perp})-(2\pi)^2\delta(k_{1\perp}) \right ]
 u(xP) \left [ \tilde U(k_\perp-k_{1\perp})-(2\pi)^2 \delta(k_\perp-k_{1\perp}) \right ]_{ba}
  \nonumber \\ &+&
 ie g^2 \int \frac{d^2 p_\perp}{(2\pi)^2}
 \frac{\rho_{p,a}(p_\perp)}{p_\perp^2} \int d^2x_\perp e^{ik_\perp \cdot x_\perp}
  \ \bar u(l_q) \varepsilon \!\!\!/ S_F(xP+q)
 \frac{n \!\!\! / }{x_gP+i\epsilon}  t^b   \left [ U(x_\perp)-1 \right ] u(xP)
  \nonumber \\ && \times
\left [ \tilde V(x_\perp)-1 \right ]_{ba}
\end{eqnarray}
The amplitude of the diagram in Fig.~5c does not receive any
contribution from the second part of the $C_{V,reg}^\mu$ term since
both $k_1^-$ poles are lying in the same half plane. One thus
obtains,
\begin{eqnarray}
{\cal M}_{5c}&=& -ie g^2 \int \frac{d k^-_1 d^2 k_{1\perp}
}{(2\pi)^3} \int \frac{d^2 p_\perp}{(2\pi)^2}
 \frac{\rho_{p,a}(p_\perp)}{p_\perp^2} \ \bar u(l_q)
\frac{C_U \!\!\!\!\!\!\!/  \ \ (q-k_1,p_\perp)}{(q-k_1)^2+i\epsilon}
t^b S_F(xP-l_\gamma+k_1)
 \varepsilon \!\!\!/
 \nonumber \\ && \times  S_F(xP+k_1) n\!\!\!/
\left [ U(k_{1\perp})-(2\pi)^2\delta(k_{1\perp}) \right ]
 u(xP) \left [ \tilde U(k_\perp-k_{1\perp})-(2\pi)^2 \delta(k_\perp-k_{1\perp}) \right ]_{ba} \ .
 \nonumber \\
\end{eqnarray}

\begin{figure}[t]
\begin{center}
\includegraphics[width=16cm]{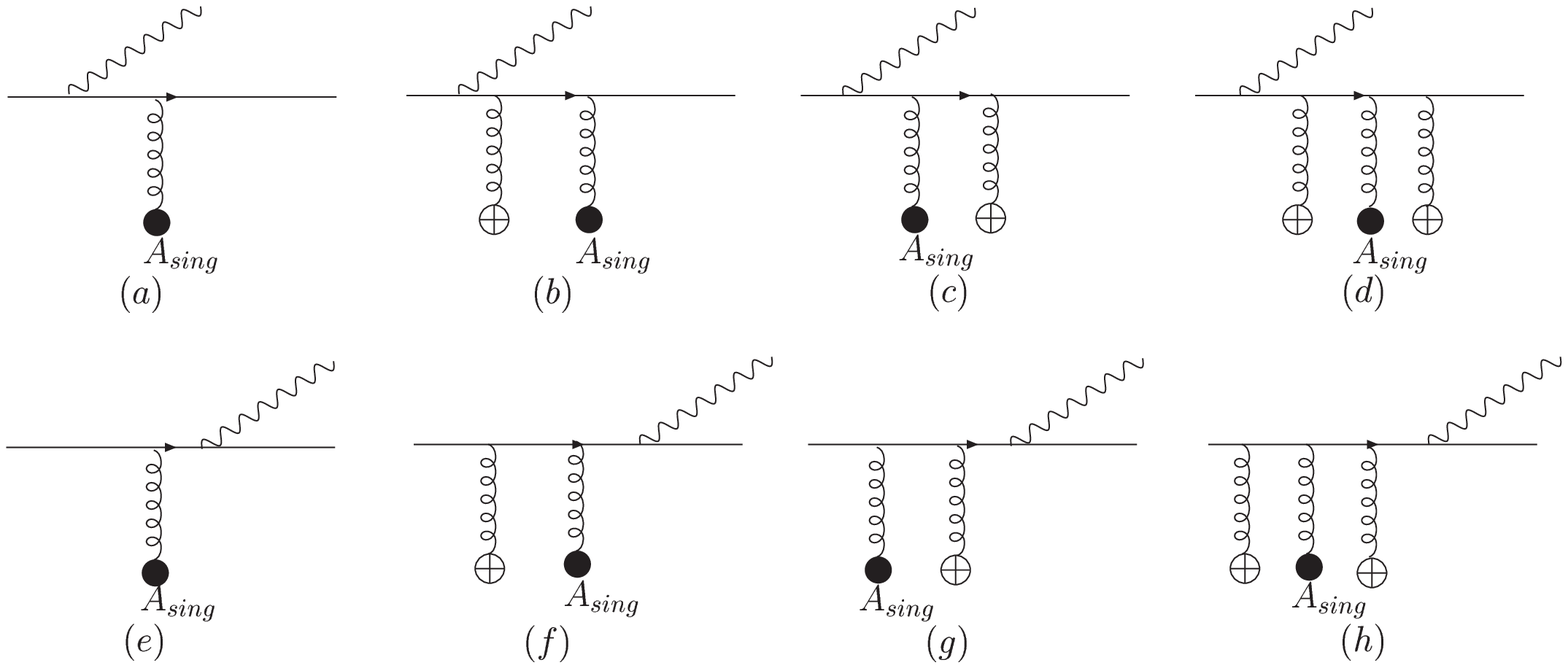}
\caption[] { The contribution from the singular terms to the spin
dependent amplitude. A black dot denotes a classical field
$A_{sing}$ insertion, while the gluon line terminated by a cross
surrounded with a circle denotes a  classical field $A_A$
insertion.} \label{6}
\end{center}
\end{figure}

We now turn to discuss the contributions from the singular terms. As
explained in Ref.~\cite{Blaizot:2004wv}, it is convenient to compute
it in coordinate space. The expression for the singular term in
coordinate space is then given by~\cite{Blaizot:2004wv},
\begin{eqnarray}
A^-_{sing}=-i \frac{g^2}{2}[ A^-_A(x) \cdot T ] \tilde
V(x^+,-\infty; x_\perp) \theta(-x^-) \frac{1}{\nabla_\perp^2}
\rho_p(x_\perp)
\end{eqnarray}
where the theta function $\theta(-x^-)$  reflects the fact that this
gluon field is created through
 an initial state interaction.
 $\tilde V(x^+,-\infty; x_\perp)$ denotes an incomplete Wilson line:
\begin{eqnarray}
\tilde V(x^+,-\infty; x_\perp)  = {\cal P} {\rm exp}
 \left [ i \frac{g}{2} \int_{-\infty}^{x^+} dz^+ A^-_A(z^+,x_\perp) \cdot T \right ]
\end{eqnarray}
In order to correctly compute the singular contributions,
 it is necessary to regularize $\delta(x^+)$ by giving it a small width
\begin{eqnarray}
\delta(x^+) \longrightarrow \delta_\epsilon(x^+)
\end{eqnarray}
where $\delta_\epsilon(x^+)$ is a regular function whose support is
$[0,\epsilon]$,
 which becomes $\delta(x^+)$ when $\epsilon$ goes to zero.  The final result is independent of the precise
 choice of the regularization. The field $A_{sing}^\mu$ is inserted on the quark line  at the
 'times' $x^+$, the incoming quark then rescatters off the field $A_A^\mu$ of the nucleus
   in the ranges $[0,x^+]$ and $[x^+,\epsilon]$. The photon can only be emitted from the quark line
 either before multiple gluon re-scattering  or after gluon re-scattering,
 because in the limit $\epsilon \rightarrow 0$,  there is not sufficient time for emitting a photon
 inside the nucleus.
The eight diagrams contributing to the amplitude are illustrated in
Fig.~\ref{6}. Combining the contributions from diagram Fig.~\ref{6}a,
Fig.~\ref{6}b, Fig.~\ref{6}c and Fig.~\ref{6}d, the resulting
amplitude in coordinate space is,
\begin{eqnarray}
{\cal M}_{6a+6b+6c+6d}&=&\int d^4 x e^{iq \cdot x} \ \bar u(l_q)
U(+\infty,x^+; x_\perp) n \!\!\!/ \left [ig t^a A_{sing}^{-a}(x)
\right ] S_F(xP-l_\gamma) \nonumber \\ && \times ie \varepsilon
\!\!\!/   U(x^+,-\infty; x_\perp) u(xP) \nonumber \\ &=&-eg \int d^4
x e^{iq \cdot x} \ \bar u(l_q) U(+\infty,x^+; x_\perp) n \!\!\!/
\left [ t^a A_{sing}^{-a}(x) \right ] S_F(xP-l_\gamma) \nonumber \\
&& \times \varepsilon \!\!\!/   U^\dag(+\infty,x^+; x_\perp)
U(+\infty,-\infty; x_\perp)  u(xP)
\end{eqnarray}
where the  incomplete Wilson lines in the fundamental representation
are defined as
\begin{eqnarray}
U(+\infty,x^+; x_\perp) &=& {\cal P } {\rm exp}
 \left [ i g \int^{+\infty}_{x^+} dz^+ A^-_A(z^+,x_\perp) \cdot t \right ]
\\
U(x^+,-\infty; x_\perp)&=&{\cal P} {\rm exp}
 \left [ i g \int_{-\infty}^{x^+} dz^+ A^-_A(z^+,x_\perp) \cdot t \right ]
\end{eqnarray}
In order to simplify this expression, we use the algebraic identity,
\begin{eqnarray}
U(+\infty,x^+; x_\perp) t^a U^\dag(+\infty,x^+; x_\perp)=t^b \tilde
U_{ba}(+\infty,x^+; x_\perp) \label{adjfun}
\end{eqnarray}
and also the formula first derived in Ref.~\cite{Blaizot:2004wv},
\begin{eqnarray}
i\frac{g}{2} \int_{-\infty}^{+\infty} dx^+ \tilde
U(+\infty,x^+;x_\perp) [A^-_A(x) \cdot T ] \tilde V(x^+,-\infty;
x_\perp)= \tilde U(x_\perp)- \tilde V(x_\perp)
\end{eqnarray}
After carrying out the $x^-$ and $x^+$ integrations, the above
expression is simplified to,
\begin{eqnarray}
{\cal M}_{6a+6b+6c+6d}&=& ieg^2 \int \frac{ d^2 p_\perp}{(2\pi)^2}
\frac{\rho_{p,a}(p_\perp)}{p_\perp^2} \int d^2 x_\perp e^{ik_\perp
\cdot x_\perp} \ \nonumber \\ && \times \bar u(l_q) n \!\!\!/ t^b
S_F(xP-l_\gamma) \varepsilon \!\!\!/    U( x_\perp)
u(xP)\frac{1}{x_gP+i\epsilon} \left [ \tilde U(x_\perp)-\tilde
V(x_\perp) \right ]_{ba}
\end{eqnarray}
Following the same procedure,
 it is straightforward to write down the amplitude from the diagrams
in Fig.~\ref{6}e, Fig.~\ref{6}f, Fig.~\ref{6}g and Fig.~\ref{6}h
\begin{eqnarray}
{\cal M}_{6e+6f+6g+6h}&=& ieg^2 \int \frac{ d^2 p_\perp}{(2\pi)^2}
\frac{\rho_{p,a}(p_\perp)}{p_\perp^2} \int d^2 x_\perp e^{ik_\perp
\cdot x_\perp} \ \nonumber \\ && \times \bar u(l_q) \varepsilon
\!\!\!/S_F(xP+q) n \!\!\!/ t^b
   U( x_\perp)  u(xP)\frac{1}{x_gP+i\epsilon} \left [ \tilde U(x_\perp)-\tilde V(x_\perp) \right ]_{ba}
\end{eqnarray}

Collecting all pieces together, the total amplitude reads,
\begin{eqnarray}
{\cal M}_{4+5+6}&=&-ie g^2  \int \frac{d^2 p_\perp}{(2\pi)^2}
 \frac{\rho_{p,a}(p_\perp)}{p_\perp^2} \int \frac{d k^-_1 d^2 k_{1\perp} }{(2\pi)^3}
 \ \bar u(l_q)
    \nonumber \\
&& \times  \left \{
  \varepsilon \!\!\!/
S_F(xP+q)
 \frac{C_U \!\!\!\!\!\!\!/  \ \ (q-k_1,p_\perp)}{(q-k_1)^2+i\epsilon}  t^b S_F(xP+k_1) n\!\!\!/U(k_{1\perp})
 \right .\
 \nonumber \\ && \  \ \ \ \ +
  \frac{C_U \!\!\!\!\!\!\!/  \ \ (q-k_1,p_\perp)}{(q-k_1)^2+i\epsilon}  t^b S_F(xP-l_\gamma+k_1)
 n\!\!\!/ U(k_{1\perp})
S_F(xP-l_\gamma) \varepsilon \!\!\!/
 \nonumber \\ &&
\left .\  \ \ \ \  + \frac{C_U \!\!\!\!\!\!\!/  \ \
(q-k_1,p_\perp)}{(q-k_1)^2+i\epsilon}  t^b S_F(xP-l_\gamma+k_1)
\varepsilon \!\!\!/
   S_F(xP+k_1) n\!\!\!/ U(k_{1\perp})
\right \}
 \nonumber \\ && \times u(xP)
 \left [ \tilde U(k_\perp-k_{1\perp})-(2\pi)^2 \delta(k_\perp-k_{1\perp}) \right ]_{ba}
  \nonumber \\
&+& ieg^2 \int \frac{ d^2 p_\perp}{(2\pi)^2}
\frac{\rho_{p,a}(p_\perp)}{p_\perp^2} \int d^2 x_\perp e^{ik_\perp
\cdot x_\perp} \ \nonumber \\ & \times& \!\!\! \bar u(l_q) \frac{  n
\!\!\!/ S_F(xP-l_\gamma) \varepsilon \!\!\!/  + \varepsilon
\!\!\!/S_F(xP+q) n \!\!\!/ }{x_gP+i\epsilon} t^b
   U( x_\perp)  u(xP)\left [ \tilde U(x_\perp)-1 \right ]_{ba}
   \label{pamp}
\end{eqnarray}
where the $\tilde V(x_\perp)$ terms drop out as expected. The spin
dependence of the total amplitude comes from the correlation between
$p_\perp$ and the transverse spin vector of the proton $S_\perp$.
With the obtained amplitude, we are ready to compute the twist-3
spin dependent cross section. However, since the calculation of the
full polarized cross section is quite involved, we restrict ourself
here to the discussion of two results, namely the derivative term
contribution in a dense medium and the cross section in the large
photon transverse momentum limit. We believe that it is sufficient
to demonstrate the most interesting feature of the complete result,
on the one hand, and to check the consistence of our formalism by
extrapolating the result to the high transverse momentum limit and
comparing it with the polarized cross section computed in the
collinear twist-3 approach, on the other hand.




\subsection{The derivative term in a dense medium }
One of the terms in the polarized cross section proportional to the
derivative of the ETQS function
 is often refereed to as the derivative term, which is usually considered to be the dominant contribution to
 the spin asymmetry  in the forward region. In this subsection, we sketch a few key steps when deriving the expression
for the derivative term.   As mentioned in the previous section, the
spin dependent hard part is calculated from an interference of two
partonic scattering amplitudes, as illustrated in Fig.~\ref{3}. We
thus  proceed by defining a hard part $H^{\mu \nu}_{Born} (
p_\perp,k_\perp)$ according to the following equation,
\begin{eqnarray}
&& \!\!\!\!\!\!\!\!\!\!\!\! \sum_{\rm spins,color} \frac{1}{2x}
{\cal M}_{4+5+6}|_{derivative}{\cal M}_3^* \delta(l_q^2) = \int
\frac{d^2 p_\perp}{(2\pi)^2} \frac{1}{x_gP+i\epsilon}
 H^{\mu \nu}_{Born} ( p_\perp,k_\perp)n_\mu n_\nu\delta(l_q^2)
 \nonumber \\ && \times
 \int d^2 x_\perp d^2 y_\perp
e^{ik_\perp \cdot ( x_\perp-y_\perp)} {\rm Tr_c} \left \{
 \left [ U^\dag(y_\perp)-1 \right ] t^b U(x_\perp)  \frac{\rho_{p,a}(p_\perp)}{p_\perp^2} \right \}
  \left [ \tilde U(x_\perp)-1 \right ]_{ba}
\end{eqnarray}
where ${\cal M}_{4+5+6}|_{derivative}$ represents the last term in
Eq.~(\ref{pamp}).
 As we shall explain below, the $C_U$ terms in Eq.~(\ref{pamp}) do not give rise to a derivative term contribution.
The next step is to expand the hard part in terms of $p_\perp$,
\begin{eqnarray}
 H^{\mu \nu}_{Born} ( p_\perp,k_\perp)\delta(l_q^2)=
 H^{\mu \nu}_{Born} ( p_\perp,k_\perp)\delta(l_q^2)|_{p_\perp=0}
 +\frac{\partial  H^{\mu \nu}_{Born} ( p_\perp,k_\perp)\delta(l_q^2)}{\partial p_\perp^\rho}|_{p_\perp=0} \ p_\perp^\rho + ...
\end{eqnarray}
where the spin dependent part is the term linear in $p_\perp$, in
which we only keep the contribution containing the derivative of the
delta function, leading to a derivative of the ETQS function by
partial integration. More precisely, the relevant contribution is
given by,
\begin{eqnarray}
 \left [ \frac{(k_\perp^\rho-l_{\gamma \perp}^\rho)}{l_q \cdot P}  H^{\mu \nu}_{Born} ( p_\perp,k_\perp)
 \frac{ \partial \delta(l_q^2)}{\partial x}\right ]_{p_\perp=0} \  p_{\perp, \rho}
\end{eqnarray}
Once the collinear expansion has been carried out, $p_\perp$ is set
to zero in the hard part as indicated in the above formula. At this
point, it becomes clear why the $C_U$ terms  do not give rise to a
derivative term contribution, namely simply because $C_U$ vanishes
when $p_\perp=0$,
\begin{eqnarray}
C_U^\mu(q-k_1,p_\perp=0)=0
\end{eqnarray}
We proceed by combining $p_\perp$ with the color source term to
yield the gluon field strength operator using Eq.~(\ref{proton}),
\begin{eqnarray}
 \frac{\rho_{p,a}(p_\perp)}{p_\perp^2}  p_{\perp, \rho} \longrightarrow F_{\rho,a}^+(p_\perp)
\end{eqnarray}
Since the hard part is independent of $p_\perp$ after the collinear
expansion, the $p_\perp$ integration can be trivially carried out.
The quark gluon correlator can subsequently be parameterized through
the ETQS function,
\begin{eqnarray}
\int \frac{d^2 p_\perp}{(2\pi)^2} \langle \bar \psi
\frac{\rho_{p,a}(p_\perp)}{p_\perp^2}  p_{\perp, \rho} \psi
\rangle_{\rm proton}
 \longrightarrow  \epsilon^{\rho S_\perp  n  p} \frac{2}{N_c^2-1}  \frac{1}{2\pi} t^a T_F(x,x+x_g)
\end{eqnarray}
where the correlation between the transverse spin vector of the
proton and $p_\perp$ becomes manifest. The color structure
associated with the ETQS function is fixed by following the argument
made in Refs.~\cite{Qiu:1991pp,Qiu:1991wg,Yuan:2009dw}. Moreover,
one needs to isolate  the imaginary part of the soft gluon pole
using the identity $\frac{1}{x_gP+i\epsilon}={\rm P}
\frac{1}{x_gP}-i\pi \delta(x_gP)$.
 The contributions from its real part cancel out between mirror diagrams.
 With all these calculation recipes, one can readily compute the contribution from the derivative term.
 The spin dependent cross section involving the derivative term takes the following form,
\begin{eqnarray}
\frac{d^3\Delta \sigma}{ d^2 l_{\gamma\perp} dz} & \propto & \int
d^2 k_\perp
 dx_g' dx \ \left [\epsilon^{l_{\gamma} S_\perp  n p}-\epsilon^{{\rm k}_{\perp} S_\perp  n p}\right ]
\frac{1}{l_q \cdot P} \left  [ H^{\mu \nu}_{Born} ( p_\perp,k_\perp)
 \frac{ \partial \delta(l_q^2)}{\partial x}\right ]_{p_\perp=0}  \sum_q e_q^2 T_{F,q}(x,x)
\nonumber \\ && \!\!\!\!\!\!\! \times \int d^2 x_\perp d^2 y_\perp
e^{ik_\perp \cdot ( x_\perp-y_\perp)}
  \langle {\rm Tr_c} \left [ \left ( U^\dag(y_\perp)-1 \right ) t^b U(x_\perp) t^a \right ] \left [ \tilde U(x_\perp)-1 \right ]_{ba} \rangle_{ x_g'}
\label{cross}
\end{eqnarray}
The expression for the soft part from the nucleus side in the above
formula can be further simplified. Using Eq.~(\ref{adjfun}), one
obtains,
\begin{eqnarray}
&& {\rm Tr_c} \left [ \left ( U^\dag(y_\perp)-1 \right ) t^b
U(x_\perp) t^a \right ] \left [ \tilde U(x_\perp)-1 \right ]_{ba}
   \nonumber \\ &&=
C_F {\rm Tr_c} \left [ U^\dag(y_\perp) U(x_\perp) \right ] -{\rm
Tr_c}\left [ U^\dag(y_\perp) t^a U(x_\perp) t^a \right ]
\label{wline}
\end{eqnarray}
Employing the Fierz identity,
\begin{eqnarray}
t^a_{ij} t^a_{kl}=\frac{1}{2} \delta_{il} \delta_{kj}-\frac{1}{2N_c}
\delta_{ij} \delta_{kl}
\end{eqnarray}
the last term in  Eq.~(\ref{wline}) is rewritten as,
\begin{eqnarray}
{\rm Tr_c}\left [ U^\dag(y_\perp) t^a U(x_\perp) t^a \right ]
=\frac{1}{2}{\rm Tr_c}\left [ U^\dag(y_\perp)\right ]{\rm Tr_c}\left
[ U(x_\perp)\right ] -\frac{1}{2N_c} {\rm Tr_c}\left [
U^\dag(y_\perp)  U(x_\perp)  \right ]
\end{eqnarray}
Inserting the above decomposition into Eq.~(\ref{wline}), we obtain,
\begin{eqnarray}
&& {\rm Tr_c} \left [ U^\dag(y_\perp) t^b U(x_\perp) t^a \right ]
\left [ \tilde U(x_\perp)-1 \right ]_{ba}
  \nonumber \\ &=&
  \frac{N_c}{2} {\rm Tr_c} \left [ U^\dag(y_\perp)  U(x_\perp) \right ]-
\frac{1}{2}{\rm Tr_c}\left [ U^\dag(y_\perp)\right ]{\rm Tr_c}\left
[ U(x_\perp)\right ]
\end{eqnarray}
where the first term can be related to the dipole type gluon
distribution, while the second term is a new contribution that
arises from the color entanglement effect.  To arrive at the final
expression for the polarized cross section, we need to explicitly
evaluate the hard part and carry out the $x_g'$ integration using
the delta function $ \delta(l_q^2)$ which originates from the on
shell condition. After combining the contributions from the left and
right cut diagrams, one ends up with,
\begin{eqnarray}
 \frac{d^3\Delta \sigma}{ d^2 l_{\gamma\perp}dz} & =&
\frac{\alpha_s \alpha_{em}  N_c}{N_c^2-1}  \frac{1+(1-z)^2}{z l_{\gamma\perp}^2}(z-1)
\int_{x_{min}}^1 dx  \int d^2 k_\perp
\frac{\left[ \epsilon^{l_{\gamma }S_\perp   n p }-\epsilon^{{\rm k}_{\perp }S_\perp   n p }\right ]}
{\left(k_\perp-l_{\gamma\perp}/z \right )^2 \left (k_\perp -l_{\gamma \perp}\right )^2}
\nonumber \\&&\times \sum_q
e_q^2 \left [ -x\frac{d}{dx}  T_{F,q}(x,x) \right ] \left [ x_g'
G_{DP}(x_g',k_\perp) - x_g'G_4(x_g', k_\perp) \right ]
\label{crosssection}
\end{eqnarray}
 which is the main result of this section.
Here we introduce a new gluon distribution  $ G_{4}(x_g', k_\perp)$.
It is defined as,
\begin{eqnarray}
x_g' G_{4}(x_g', k_\perp)&=& \frac{k_\perp^2 N_c}{2 \pi^2 \alpha_s}
\int \frac{d^2x_\perp d^2y_\perp }{(2\pi)^2} e^{i  k_\perp \cdot
(x_\perp-y_\perp)} \frac{1}{N_c^2} \langle {\rm Tr_c} [U(x_\perp)]
{\rm Tr_c}[ U^\dag(y_\perp)] \rangle_{x_g'} \label{g4}
\end{eqnarray}
This new gluon distribution only shows up in the spin dependent
cross section and is absent in the unpolarized cross section. The
extra gluon exchange between the remnant of the proton and active
partons plays a crucial role in yielding the nontrivial Wilson line
structure in the Eq.~(\ref{g4}). The additional term  associated
with $G_{4}(x_g', k_\perp)$ thus essentially arises from the color
entanglement effect.

More interestingly, the gluon distribution $G_{4}(x_g', k_\perp)$
can be calculated in the MV model through a recursion procedure
systemically developed in Ref.~\cite{Blaizot:2004wv}. In the MV
model, to our surprise, it is simply given by,
\begin{eqnarray}
x_g' G_{4}(x_g', k_\perp)&=&\frac{k_\perp^2 N_c}{2 \pi^2 \alpha_s} \pi
R_0^2 \int \frac{d^2r_\perp }{(2\pi)^2} e^{i  k_\perp \cdot r_\perp}
\frac{1}{N_c^2}   e^{-\frac{1}{4} r_\perp^2 Q_{sq}^2}
\nonumber \\
 &=&\frac{1}{N_c^2}
G_{DP}(x_g', k_\perp)
\label{g4}
\end{eqnarray}
We thus conclude that the novel gluon distribution $G_4$ is sizable, though
it is suppressed in the large $N_c$ limit as compared to the dipole type gluon distribution.

We now make some observations on the polarized cross section in the
different kinematic limits. At small photon transverse momentum
$\Lambda_{QCD} \ll l_{\gamma \perp} \ll Q_{sq}\sim k_\perp$, the
denominator in the Eq.~(\ref{crosssection}) can be approximated as,
\begin{eqnarray}
\frac{1}{\left (k_\perp-l_{\gamma\perp}/z \right )^2 \left ( k_\perp-l_{\gamma \perp} \right )^2}\approx
\left( 1+\frac{1+z}{z} \frac{2k_\perp \cdot l_{\gamma \perp}}{ k_\perp^2} \right )\frac{ 1}{k_\perp^4}
\end{eqnarray}
 Once adopting
such approximation in the both unpolarized cross section and polarized cross section,
it is easy to see that the spin asymmetry computed in the hybrid approach scales as $l_\perp$ at low transverse momentum.
As a comparison, in the standard collinear twist-3
framework, the predicated spin asymmetry is proportional to $ 1/l_\perp$.
For $ l_{\gamma\perp} \gg Q_{sq}\sim k_\perp$, the denominator can
be approximated by $1/(k_\perp-l_{\gamma\perp}/z)^2\approx
z^2/l_{\gamma\perp}^2$. Using  Eq.~\ref{relation1} and Eq.~\ref{g4},
 the polarized cross section is correspondingly reduced to,
\begin{eqnarray}
 \frac{d^3\Delta \sigma}{ d^2 l_{\gamma\perp}dz} & =&
\frac{\alpha_s \alpha_{em}  N_c}{N_c^2-1} \epsilon^{l_{\gamma }
S_\perp  n p } \frac{z [1+(1-z)^2]}{l_{\gamma\perp}^6}(z-1)
\nonumber \\ && \times \sum_q e_q^2 \int_{x_{min}}^1 dx
\left [ -x\frac{d}{dx} T_{F,q}(x,x) \right ]
\left \{  x_g' G(x_g')- \frac{1}{N_c^2}  x_g' G(x_g') \right \}
\end{eqnarray}
which recovers the result for the derivative term contribution
computed in the collinear approach if one ignores the second term of the soft parts that arises from
color entanglement effect. In the next
subsection, we show that the non-derivative term contribution also
can be reproduced in our hybrid formalism in the kinematical limit
where $ l_{\gamma\perp} \gg Q_{sq}\sim k_\perp$ provided that the $G_4$ contribution is neglected.

We now close this subsection with a few further remarks.
\begin{itemize}
\item
The color entanglement effect discovered for double spin asymmetries
(DSA) leads to a violation of generalized TMD
factorization~\cite{Rogers:2010dm}. In contrast, the process we
study is factorizable though there exists an additional term arises
from the color entanglement effect. The reason is that collinear
factorization is applied on the proton side and the basic building
block of the soft part on the nucleus side, namely the Wilson line,
is a universal object.
\item
Apparently, the gluon distribution $G_{4}(x_g', k_\perp)$  vanishes
in the single gluon exchange approximation, thus requireing at least
two gluon exchange. This is in line with the argument that two extra
gluon attachments from an unpolarized target are required to
generate a non-trivial color entanglement
effect~\cite{Rogers:2010dm}.
\item
Since $G_{4}(x_g', k_\perp)$ can be explicitly evaluated in the MV
model, one can test it by measuring  SSAs for photon production in
p$^{\uparrow}$A collisions.
\item
In general color entanglement plays a less important role for
p$^{\uparrow}$p collisions than for p$^{\uparrow}$A collisions as
the existence of $G_{4}(x_g', k_\perp)$ requires at least two gluons exchange. However,
the SSA for photon production in p$^{\uparrow}$p collisions might
receive the significant contribution from the additional term
proportional to the distribution $G_{4}(x_g', k_\perp)$ in the very
forward region.
\item
If we apply collinear factorization on both the proton and nucleus
sides, according to the model calculation result Eq.~\ref{g4}, the color entanglement effect would survive.
This might indicate that the collinear higher-twist factorization breaks down in the process we study.
\end{itemize}

\subsection{The collinear limit}
In this subsection, we show that not only the derivative term
contribution but also the non-derivative term contribution computed
in the collinear factorization framework can be recovered from our
hybrid approach in the limit $ l_{\gamma\perp} \gg Q_{sq}$ after neglecting the additional contribution results  from
color entanglement effect. To achieve this goal, our main strategy is to systematically neglect
all terms suppressed by powers of
 $Q_{sq}/l_{\gamma\perp}$(or  $k_\perp/l_{\gamma\perp}$).

The first step is to set $k_{1\perp}=0$ in the hard part in
Eq.~(\ref{pamp}). This is a well justified approximation in the
kinematical limit that we consider because the typical transverse
momentum carried by small x gluons is of the order of $Q_{sq}$ and
thus much smaller than the photon transverse momentum $
l_{\gamma\perp}$. We then can trivially carry out the $k_{1\perp}$
and $k^-_1$ integrations in Eq.~(\ref{pamp}) after applying the
  Eikonal approximation to the quark propagators $S_F(xP+k_1)$ and $S_F(xP-l_\gamma+k_1)$. As a result,
the polarized amplitude simplifies to,
\begin{eqnarray}
&&\!\!\!\!\!\! {\cal M}_{4+5+6}\approx  - ieg^2 \int \frac{ d^2
p_\perp}{(2\pi)^2} \frac{\rho_{p,a}(p_\perp)}{p_\perp^2} \int d^2
x_\perp e^{ik_\perp \cdot x_\perp} \ \bar u(l_q) \nonumber \\ & &
\times
 \frac{  C_L \!\!\!\!\!\!\!/   \ \ (q,p_\perp)
S_F(xP-l_\gamma) \varepsilon \!\!\!/  + \varepsilon \!\!\!/S_F(xP+q)
C_L \!\!\!\!\!\!\!/ \ \ (q,p_\perp)  }{q^2+i\epsilon} t^b
 U( x_\perp)  u(xP)\left [ \tilde U(x_\perp)-1 \right ]_{ba}
\end{eqnarray}
where $C_L \!\!\!\!\!\!\! /  \ \ $ is the well known effective
Lipatov vertex for the production of a gluon via the fusion of two
gluons. It is given by,
\begin{eqnarray}
\frac{C_L \!\!\!\!\!\!\!/ \ \ (q,p_\perp)}{q^2+i\epsilon}
&=&\frac{C_U \!\!\!\!\!\!\!/ \ \ (q,p_\perp)}{q^2+i\epsilon}-\frac{n
\!\!\!/}{q^++i\epsilon} \nonumber \\&=&
\frac{-p_\perp^2}{(q^2+i\epsilon)(q^-+i\epsilon)}
p\!\!\!/+\frac{k_\perp^2}{(q^2+i\epsilon)(q^++i\epsilon)} n\!\!\!/
-2 \frac{q^- n\!\!\!/+{\rm p}_\perp \!\!\!\!\!\!/}{q^2+i\epsilon}
\end{eqnarray}
One notices that the first term in the above formula can be
neglected since it is beyond the order in $p_\perp$ that we
consider. The second term contains one soft gluon pole and one hard
gluon pole,
\begin{eqnarray}
\frac{1}{q^++i\epsilon} \longrightarrow  x_g=0 , \ \ \ \ \ \ \
\frac{1}{q^2+i\epsilon} \longrightarrow
x_g=\frac{(k_\perp+p_\perp)^2}{2x_g' P \cdot \bar P}
\end{eqnarray}
When the photon transverse momentum is much larger than the
saturation scale,
 the hard gluon pole $x_g=(k_\perp+p_\perp)^2/(2x_g' P \cdot \bar P)\approx 0$
degenerates with the soft gluon pole. It is easy to verify that the
contributions from both cancel in this kinematical limit. We are
thus left with the last term. Using the Ward identity, it can be
replaced with
\begin{eqnarray}
-2 \frac{q^- n\!\!\!/+{\rm p}_\perp \!\!\!\!\!\!/}{q^2+i\epsilon}
\longrightarrow 2 \frac{q^+ p\!\!\!/+{\rm k}_\perp
\!\!\!\!\!\!/}{q^2+i\epsilon}
\end{eqnarray}
Applying the Ward identity to the right side of the cut diagrams, we
may make the following replacement for the gluon polarization
vector,
\begin{eqnarray}
n^\mu \longrightarrow \frac{-{\rm k}_\perp^\mu }{q^-}
\end{eqnarray}
With these replacements, the hard part can be written as,
\begin{eqnarray}
-\frac{1}{q^-}\delta(q^2) ( x_gP_\mu+{\rm k}_{\perp,\mu}){\rm k}_{\perp,\nu}
H^{\mu \nu}_{Born} ( p_\perp,k_\perp)\delta(l_q^2)
\end{eqnarray}
where $H^{\mu \nu}_{Born} ( p_\perp,k_\perp)$ has been defined in
the previous subsection. To proceed further, we keep the leading
term  which is proportional to $k_\perp^2$ and neglect all higher
order terms in $k_\perp$.
 After averaging over the azimuthal angle of $k_\perp$, the hard part reads,
\begin{eqnarray}
 \frac{1}{2q^-} k_\perp^2 \left \{ \frac{  P_\mu {\rm p}_{ \perp,\nu}}{x_g'P\cdot \bar P}
 + d_{\perp,\mu \nu} \right \}
 \left [ H^{\mu \nu}_{Born} ( p_\perp,k_\perp)\delta(l_q^2)\delta(q^2) \right ]_{k_\perp=0}
 \label{hard}
\end{eqnarray}
where the tensor $ d_{\perp,\mu \nu}$ is defined as
 $ d_{\perp,\mu \nu}=-g_{\mu \nu}+(p_\mu n_\nu+p_\nu n_\mu)/p \cdot n$.
$k_\perp^2$ in the above formula can be combined with the soft part,
namely the Wilson lines, and related to the transverse momentum
dependent gluon distributions $G_{DP}(x_g',k_\perp)$ and
$G_{4}(x_g', k_\perp)$. Since there is no  $k_\perp$ dependence in
the hard part any longer,
 the $k_\perp$ integration can be trivially carried out using the relations
 presented in Eq.~(\ref{relation1}) and Eq.~(\ref{g4}). The resulting soft part is simply the
ordinary integrated gluon distribution if we ignore the term generated from $G_4$.

On the other hand, the spin dependent hard part from Fig.~\ref{1}a
calculated in the collinear approach is proportional to the
following expression
\begin{eqnarray}
\frac{1}{2} d_{\perp,\nu }^\sigma  \left [ H^{\mu \nu}_{Born} (
p_\perp,k_\perp)\delta(l_q^2)\delta(q^2) \right ]_{k_\perp=0}
\Lambda_{\sigma  \rho \mu} p^\rho \label{fig1aH}
\end{eqnarray}
with the three gluon vertex being defined as,
\begin{eqnarray}
\Lambda_{\sigma \rho \mu }&=&g_{\sigma \rho}(x_g'\bar P-x_gP-{\rm
p_\perp})_\mu +g_{\rho \mu}(2x_gP+2{\rm p_\perp}+x_g' \bar P)_\sigma
\nonumber \\
&+&g_{\mu \sigma}(-2x_g' \bar P-x_g P-{\rm p_\perp})_\rho
\end{eqnarray}
After few algebra steps, one finds that the expressions~(\ref{hard})
and~(\ref{fig1aH}) for the hard parts derived in the two different formalisms agree
with each other up to some trivial pre-factor. Following the
procedure outlined above, one can show that both formalisms also
yield the same hard parts for the mirror diagrams. Therefore, we
confirmed that our result without the $G_4$ related contribution being included reduces
to that computed in the standard collinear approach in the
kinematical region $l_{\gamma\perp} \gg Q_{sq}$. We consider this as
an important consistency check for the hybrid approach.

\section{Summary}
We studied the SSA in direct photon production in polarized
p$^{\uparrow}$A collisions. The calculation is carried out using a
hybrid approach in which the nucleus is treated in the CGC framework
while the collinear twist-3 formalism is applied on the proton side.
We derived the part of the polarized cross section containing the
derivative term,
 with particular emphasis on the contribution caused by the color entanglement effect.
This effect arises from the non-trivial
 interplay between one extra gluon exchange from the proton side and multiple gluon exchanges from the nucleus.
The identified new gluon distribution $G_4(x_g',k_\perp)$ can be
explicitly
 evaluated in the MV model. As a result, measuring this observable would provide us
with a unique chance to quantitatively study color entanglement
effects.

We have further shown that the spin dependent cross section computed
in the standard collinear approach can be recovered from the hybrid
approach in the kinematical region where the transverse momentum of
the produced photon is much larger than the saturation momentum,
provided that the contribution arises from color entanglement effect is not included in our result.
However, at low photon transverse momentum, in sharp contrast to the predication from the standard collinear twist-3 approach,
 the spin asymmetry is  found to be proportional to  the photon transverse momentum.

A direct extension of this work is to investigate the impact of the
color entanglement effect on the SSA in Drell-Yan lepton pair
production in p$^{\uparrow}$A collisions. One can also use the
hybrid formalism to calculate the SSAs for pion production and
di-jet and photon-jet~\cite{Schafer:2014xpa} production in p$^{\uparrow}$A collisions. A proposed
p$^{\uparrow}$A program at RHIC~\cite{Aschenauer:2013woa}  is thus
extremely welcome. Finally, we would like to emphasize that the
hybrid approach in principal can also be applied to p$^{\uparrow}$p
collisions, where it, however, only is valid in the very forward region at low transverse momentum
$\Lambda_{QCD} \ll l_{\gamma\perp} \leq Q_{sq}^p$ (where $Q_{sq}^p$
is the proton saturation momentum).
 In other words, the standard collinear twist-3 approach is not adequate to describe SSA phenomenology
in p$^{\uparrow}$p or p$^{\uparrow}$A collisions
in the mentioned kinematic region where the CGC framework can apply on target nucleus or proton side.

\vskip 1 cm {\bf Acknowledgments:} We are grateful to Bowen Xiao and Feng Yuan for
reminding us of the tadpole type contribution to the gluon distribution $G_4$ in the MV model.
This work has been supported by BMBF (OR 06RY9191).

\end {document}